\documentclass[11pt]{article}

\usepackage[top=50mm, bottom=50mm, left=50mm, right=50mm]{geometry}
\usepackage{lineno}
\usepackage{amssymb}
\usepackage{amsmath}
\usepackage{amsthm}
\usepackage{epsfig}
\usepackage{graphicx}
\usepackage{graphics}
\usepackage{float}
\usepackage{subcaption}
\usepackage{multirow}
\usepackage{color}
\usepackage{lineno}
\usepackage{fullpage}
\usepackage[normalem]{ulem} 
\usepackage{makeidx}
\usepackage{xspace}
\usepackage{wrapfig}
\usepackage{xcolor}
\usepackage{url}
\usepackage{authblk}

\definecolor{darkgreen}{rgb}{0.0, 0.2, 0.13}

\DeclareMathOperator*{\argmin}{arg\,min}

\newcommand\REV[1]{{\color{black} #1}}

\title{{\tt SUIHTER}: A new mathematical model for COVID-19. Application {to the analysis of the second} epidemic outbreak in Italy}
\author[1]{Nicola Parolini}
\author[1]{Luca Dede'}
\author[1]{Paola F. Antonietti}
\author[1]{Giovanni Ardenghi}
\author[1]{Andrea Manzoni}
\author[1]{Edie Miglio}
\author[2]{Andrea Pugliese}
\author[1]{Marco Verani}
\author[1,3]{Alfio Quarteroni}
\affil[1]{MOX, Department of Mathematics, Politecnico di Milano, Italy}
\affil[2]{Department of Mathematics, University of Trento, Italy}
\affil[3]{Institute of Mathematics, \'Ecole Polytechnique F\'ed\'erale de Lausanne
(EPFL), Switzerland (Professor Emeritus)}

\begin{document}

\bibliographystyle{plain}

\maketitle{}
\setcounter{page}{1}

\begin{abstract}
The COVID-19 epidemic is the last of a long list of pandemics that have affected {humankind} in the last century. In this paper, we propose a novel mathematical epidemiological model named {\tt SUIHTER} {from the names} of the seven compartments that it comprises: susceptible uninfected individuals (S), undetected (both asymptomatic and symptomatic) infected (U), isolated \REV{infected} (I), hospitalized (H),
 threatened (T), extinct (E), and recovered (R). A suitable parameter calibration that is based on the combined use of least squares method and {Markov Chain Monte Carlo (MCMC)} method is proposed with the aim of reproducing the past history of the epidemic {in Italy, surfaced in late February and still ongoing to date,} and of {validating {\tt SUIHTER} in terms of its predicting capabilities}. {A distinctive feature of the new model is that it allows a one-to-one calibration strategy between the model compartments and {the data that are daily} made available from the Italian Civil Protection.}  The new model is then applied to the {analysis of the Italian epidemic with emphasis on the second outbreak emerged in Fall 2020}. {In particular, we show that the epidemiological model {\tt SUIHTER} can be suitably used in a predictive manner to perform scenario analysis at national level}.
\end{abstract}

\section{Introduction}

{The Coronavirus pandemic of coronavirus disease 2019 (COVID-19) is a tremendous threat to global health.} Since the outbreak in early December 2019 in  China, more than {$1\, 834 \,573$ global deaths have been registered}, while the estimated total number of confirmed cases is {$84 \, 511\, 153$ up to January 2nd, 2021} \cite{Dashboard}. The real number of people infected is unknown, but probably much higher. 

In this scenario, predicting the trend of the epidemic is of paramount importance to mitigate the pressure on the health systems and activate control strategies (e.g. quarantines, lock-downs,
and suspension of travel) aiming at containing the disease and delaying the spread. 

As these predictions have vital consequences on the different actions taken from governments to limit and control the
COVID-19 {pandemic}, the recent period has seen {considerable flowering} of epidemiological mathematical models; see, e.g, \cite{Bertuzzo,DellaRossa,Gatto,sidarthe,LoliPiccolominiZama_2020,PEIRLINCK2020113410}. However, estimates and scenarios emerging from modeling highly depend on different factors, ranging from epidemiological assumptions to, perhaps most importantly, the completeness and quality of the data {based on} which models are calibrated. Since the beginning of the COVID-19 emergency, the quality of data on infections, deaths, tests, and other factors have been spoiled by under-detection
or inconsistent detection of cases, reporting delays, and poor documentation. 
This inconvenient {has affected, and still is to date hampering,} the intrinsic predictive capability of  mathematical models. 

{Despite the lack or incompleteness of the available data, which makes modeling the current COVID-19 outbreak challenging, mathematical models are still vital to establish predictions within reasonable ranges, and can be adapted to incorporate the effects of public health authority interventions in order to estimate in advance their effectiveness and their impact on the COVID-19 spread.
}
Building upon the celebrated SIR (susceptible (S), infectious (I), and  recovered (R)) model proposed in 1927 by Kermack and McKendrick \cite{kermack}, several generalizations have been formulated over the years by enriching the number of compartments, {e.g. Susceptible – Exposed – Infectious – Recovered (SEIR), Susceptible - Infectious - Susceptible (SIS), Susceptible - Exposed - Infected - Recovered - Deceased (SEIRD), Susceptible – Exposed – Infectious – Asymptomatic – Recovered (SEIAR), Susceptible - Infectious - Susceptible - Recovered (SIRS), Susceptible - Exposed - Infectious - Quarantined - Recovered (SEIQR), Maternally - derived immunity - Susceptible – Exposed – Infectious – Recovered (MSEIR), ... }; we refer to, e.g., \cite{brauer,Hethcote,martcheva} for an overview. Overall, these models have been abundantly applied to locally analyze COVID-19 outbreak dynamics in various countries (see, e.g., \cite{KUCHARSKI,Li489,LoliPiccolominiZama_2020,Maier}).

However, the peculiar epidemiological traits of the COVID-19 ask for {models better} able to accurately portray the mutable dynamic characteristics of the ongoing epidemic, with particular emphasis on two critical aspects:  (i) the crucial role played by the undetected (both asymptomatic and symptomatic) individuals; (ii) the number of individuals that require Intensive Care Unit (ICU) admission. This latter aspect is of paramount importance in designing realistic scenarios that incorporate the pressure of the epidemic on the national health systems.  

{
In this paper we introduce a new mathematical model, named {{\tt SUIHTER}}, based on the initials of the seven compartments that it comprises: susceptible uninfected individuals (S), undetected (both asymptomatic and symptomatic) infected (U), isolated \REV{infected} (I), hospitalized (H),
 threatened (T), extinct (E), recovered (R).
}
It is a system of {coupled} {ordinary differential equations (ODEs)} that are 
{driven by a set of parameters that are indeed piecewise constant time dependent functions.} A first set of parameters denote the transmission rates due to contacts between {susceptible} and undetected, quarantined or hospitalized subjects. A second set {of parameters} mimics the rates at which {I (isolated) and H (hospitalized)} individuals develop clinically relevant or life-threatening symptoms. A further parameter indicates the probability rate of detection of {previously} {undetected} infected individuals. Another set {of parameters} indicates the rate of recovery for the four classes of infected subjects. Finally, the last parameters denote the mortality rates for the different compartments.

{This {\tt SUIHTER} model has been conceived to face some of the limitations that can be found in existing epidemiological models applied to the COVID-19 pandemic. On the one hand, {some} studies adopt simple {SIR-like models \cite{KUCHARSKI,LoliPiccolominiZama_2020,Maier},} which have the advantage of having a limited number of parameters to be calibrated, but pay the price {of being unable to track} 
the dynamics of different categories of infected individuals. On the other hand, other multi-compartmental models (see e.g. \cite{Bertuzzo,sidarthe}) have been proposed to account for the detailed knowledge of the clinical characterization for different classes of infected individuals according to the actual level of disease severity. However, it is not always possible (and, even when possible, it is not easy) to associate the multiple infected compartments to the available data. 
The {\tt SUIHTER} model has been designed with the objective of creating the most compact model able to predict the different categories of infectious individuals which are considered relevant by the policy makers. 

\REV{
A key challenge in modelling the dynamics of COVID-19 epidemic is represented by the large number of the undetected cases. Indeed, the contribution to the spread of the epidemics due to the (often asymptomatic) undetected cases is too relevant to be neglected. Several authors \cite{sidarthe,10.1371/journal.pone.0240649,LIU2021110501}  have attempted, using different strategies, to quantify the number of undetected infections and their effect on epidemic spread.
In the present work, we propose a strategy about the initialization of those compartments that are not covered by the data (such as \textit{Susceptible}, \textit{Undetected} and \textit{Recovered}).}

The model adopts a two-step calibration process based on a preliminary  {estimation} of the model parameters {that uses} a Least Squares minimization, {followed by a Bayesian calibration performed through} 
{a Markov Chain Monte Carlo algorithm.} 

The model has been adopted to simulate the second COVID-19 epidemic outbreak in Italy arisen in {Fall 2020} (and still ongoing). \REV{In particular, we have investigated the  capability of the model in forecasting the occurrence of a peak for the most relevant compartments with an adequate advance notice. Results of the calibration, simulation by {\tt SUIHTER} and predictions for Italy and the six largest Italian regions are also reported.}}

The outline of the paper is as follows: in Section 2 we introduce the {{{\tt SUIHTER}}} mathematical model; Section 3 is devoted to the description of the calibration procedure, {Section 4 contains the numerical results along with their discussion.} {In Section \ref{sec:conclusions}, we draw our conclusions and we discuss some model’s limitations.}

\section{Mathematical model}

\begin{figure}[t]
\centering
\includegraphics[width=0.8\textwidth]{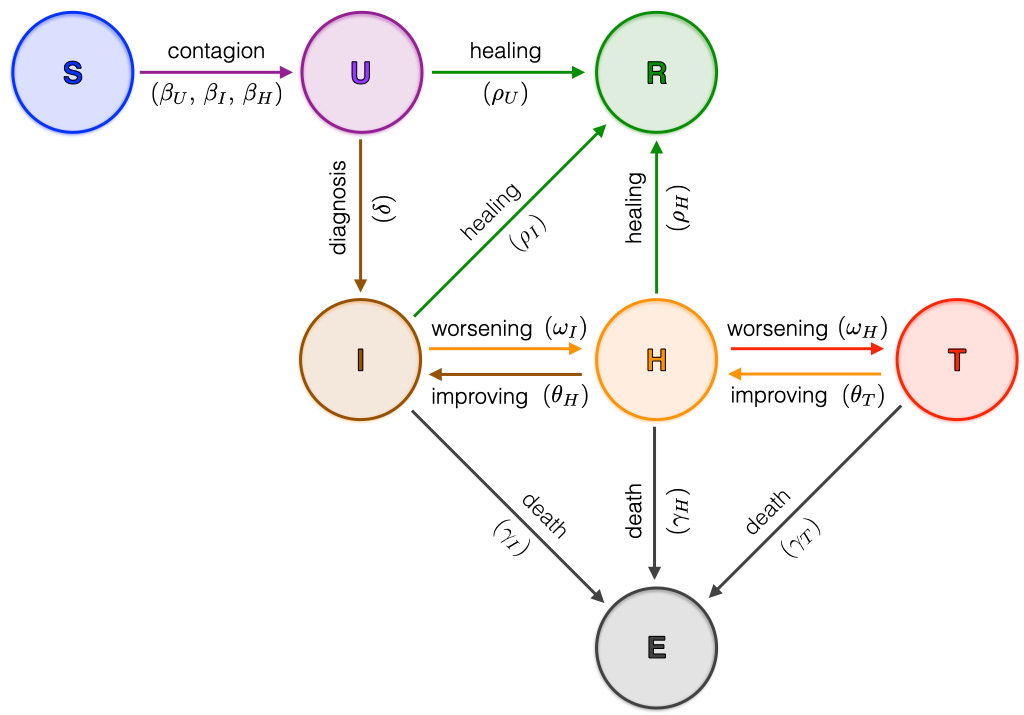}
\caption{Interactions among compartments in {\tt SUIHTER} model }
\label{fig:sidarthe}
\end{figure}

The spread of COVID-19 had made it clear that it is of paramount importance to include in epidemiological models a compartment describing the dynamics of infected  individuals that are still undetected. This is, e.g., the case of \cite{sidarthe}. However, some compartments presented in \cite{sidarthe}  (undetected asymptomatic infected and undetected symptomatic infected) are virtually impossible to be validated since {these classes of individuals cannot be traced in public databases}  (cf. \cite{PCM-DPC}). For this reason, building upon \cite{sidarthe} we propose a new model more suited to taking full advantage of publicly available data. In particular, our model is described by the following system of ordinary differential equations

\begin{equation}\label{eq:suihter}
\begin{array}{l}
\displaystyle \dot S\left( t \right) = - S\left( t \right)\frac{\beta_U U\left( t \right) + \beta_I I\left( t \right) + \beta_H H\left( t \right)}{N}, \\[3mm]
\displaystyle \dot U\left( t \right) = S\left( t \right)\frac{\beta_U U\left( t \right) + \beta_I I\left( t \right) + \beta_H H\left( t \right)}{N} - \left( {\delta + \rho_U } \right)U\left( t \right), \\[3mm]
\dot I\left( t \right) = \delta U\left( t \right) - \left( {\rho_I + \omega_I {+\gamma_I}} \right)I\left( t \right) {+ \theta_H H\left( t \right)},\\[3mm]
\dot H\left( t \right) = \omega_I I \left( t \right) - (\rho_H + \omega_H {+ \theta_H} {+\gamma_H}) H(t) { + \theta_T T\left( t \right)}, \\[3mm]
\REV{\dot T\left( t \right) = \omega_H H \left( t \right)  - \left( {{  \theta_T + \gamma_T} } \right)T\left( t \right),} \\[3mm]
  \dot E\left( t \right) = { \gamma_I I\left( t \right)} { + \gamma_H H\left( t \right)}  { +\gamma_T} T\left( t \right), \\[3mm]
  \REV{\dot R\left( t \right) =  \rho_U U\left( t \right) + \rho_I I\left( t \right) + \rho_H H\left( t \right)}
\end{array}
\end{equation}
where the compartments of the model are defined as {follows} (see Figure \ref{fig:sidarthe}):
\REV{
\begin{itemize}
\item $S$: number of \textit{susceptible} (uninfected) individuals;
\item $U$: number of \textit{undetected} (both asymptomatic and symptomatic) infected individuals;
\item $I$: number of infected individuals \textit{isolated} at home;
\item $H$: number of infected \textit{hospitalized} individuals;
\item $T$: number of infected \textit{threatened} individuals hosted in ICUs;
\item $E$: number of \textit{extinct} individuals;
\item $R$: number of \textit{recovered} individuals,
\end{itemize}
}
\noindent and $N=S+U+I+H+T+E+R$ denotes the total population (assumed constant).

The model is characterized by the following {$14$ parameters}, {some of which are possibly chosen as time dependent piecewise polynomial functions}:
\begin{itemize}
\item $\beta_U$, $\beta_I$, $\beta_H$ denote the transmission rates due to contacts between a susceptible subject and an undetected infected, a quarantined, or a hospitalized subject, respectively;
\item $\omega_I$ denotes the rate at which $I$-individuals develop clinically relevant symptoms, while $\omega_H$
denotes the rate at which $H$-individuals develop life-threatening symptoms;
{\item $\theta_H$ and $\theta_T$} {denote the rates at which $H$ and $T$-individuals improve their health conditions and return to the less critical $I$ and $H$ compartments, respectively;}
\item $\delta$ denotes the probability rate of detection, relative to undetected infected individuals;
\REV{\item $\rho_U$, $\rho_I$ and $\rho_H$ denote the rate of recovery for three classes ($U$, $I$ and $H$, respectively) of infected subjects;}
\item {$\gamma_I$, $\gamma_H$ and $\gamma_T$ denote the mortality rates for the individuals isolated at home, hospitalized and hosted in ICUs, respectively.}
\end{itemize}

\REV{Since data available for the \textit{Recovered} cases do not include those individuals who recovered before being detected, {we also propose a novel indicator, that will be denoted as \textit{Recovered from detected}, and that we define as 
$$R_D(t)=\int_{t_I}^t \left(\rho_I I(\tau)+\rho_H H(\tau)\right)\,d\tau.$$
This indicator can be obtained in postprocessing from computed compartments and collects those individuals who recovered after being detected.}}

In mathematical epidemiology a fundamental quantity is the {\it basic reproduction number} (denoted by $\mathcal{R}_0$), which is used to measure the transmission potential of a disease. It represents the average number of secondary infections produced by a typical case of an infection in a population where everyone is susceptible (see \cite{DHbook,brauer,martcheva}). For our model, { by using a similar argument to the one adopted in the proof of Proposition 1 in \cite{sidarthe}, we find
\begin{equation}\label{eq:r0}
\mathcal{R}_0 = \frac{\beta_U}{r_1 } + \frac{\delta}{ r_1}\left( \frac{\beta_I (r_3 r_4 - \theta_T\omega_H) + \beta_H \omega_I r_4}
{r_2 r_3 r_4 - r_4 \theta_H \omega_I - r_2 \theta_T \omega_H }\right),
\end{equation}
where $r_1= \delta + \rho_U, r_2=\rho_I+\omega_I + \gamma_I, r_3=\rho_H + \omega_H + \theta_H + \gamma_H, $ and $r_4=\theta_T+\gamma_T$.}
For the sake of comparison (cf. Eq.~$(32)$ in \cite{sidarthe}), we observe that in the present context the characteristic polynomial $q(s)$ of the Jacobian matrix associated to the linearization of \eqref{eq:suihter} around the equilibrium configuration $(\bar{S},0,0,0,0,\bar{E},\bar{R})$ with $\bar{S}+\bar{E}+\bar{R}=N$ is 
$$q(s)=s^3p(s)\qquad\text{with}\quad p(s)=D(s)-\bar{S} N(s)$$
where

$$
D(s)=(s+r_1)(s+r_2)(s+r_3)(s+r_4)
-(s+r_1)\theta_H\omega_I - (s+r_1)(s+r_2)\theta_T \omega_H
$$
and
$$
N(s)= (s+r_4)\{\beta_U[(s+r_2)(s+r_3)-\omega_I\theta_H]+\beta_I\delta(s+r_3)+\beta_H\delta\omega_I\}-\beta_U\omega_H\theta_T(s+r_2)-\beta_I \delta \theta_T \omega_H.
$$

From the mathematical point of view, the reproduction number $\mathcal{R}_0$ plays the role of a threshold value at the outset of the epidemic. If $\mathcal{R}_0 > 1$, the disease spreads in the population; if $\mathcal{R}_0 < 1$, the number of infected gradually declines to zero. {Note that all factor in Eq.~\eqref{eq:r0} are, as expected, actually positive. Furthermore, the expression~\eqref{eq:r0} considerably simplifies upon assuming, as done in the remainder of this paper, that \REV{$\beta_I=\beta_H=\theta_H=\gamma_H = 0$}.}

{Our {\tt SUIHTER} model, as other compartmental models, corresponds to a particular case of an integral model with arbitrary distribution of infectious time, for which $\mathcal{R}_0$ is well-known \cite{DHbook}.}

\REV{
\subsection{Model initialization}\label{sec:initialization}
A critical issue is the way those compartments for which data are unavailable (\textit{Susceptible}, \textit{Undetected} and \textit{Recovered}) are initialized.
In particular, when the analysis focuses on a late phase of the epidemics, as in the present investigation of the second epidemic outbreak in the Fall 2020, it may difficult to estimate those "initial" values as a result of the simulation from day $0$.

For these reasons, we have devised a strategy to estimate the number of \textit{Recovered} and \textit{Undetected} individuals based on the value of the \textit{Infection Fatality Ratio} (IFR) defined as the ratio between the number of deaths and the number of resolved cases (dead or recovered) at a specific time (ideally at the end of the epidemic) 
\begin{equation}\label{eq:IFR}
    \text{IFR}=\frac{E}{R+E}.
\end{equation}
We assume that IFR will be roughly constant in time, at least over the first wave.
By using the age-dependent estimates given in \cite{imperialreport} the IFR can be estimated around $1.2\, \% $ for Italy. The number of  \textit{Recovered} individuals on a given day can be then computed from \eqref{eq:IFR} based on IFR and the number of \textit{Expired} individuals. 
Moreover, the number of \textit{Undetected} individuals
 at a given time can be obtained by exploiting the detecting ratio at any given time as follows. We introduce a time dependent \textit{Case Fatality Ratio} (CFR) defined as:
\begin{equation}\label{eq:CFR}
    \text{CFR}(t)=\frac{\Delta E(t)}{\Delta R_D(t)+\Delta E(t)},
\end{equation}
where $\Delta E(t)=E(t+\Delta t/2)-E(t-\Delta t/2)$ and $\Delta R_D(t)=R_D(t+\Delta t/2)-R_D(t-\Delta t/2)$, denote  the deaths and (detected) recovered cases observed in a time-window of size $\Delta t=28$ days around a given time $t$.
The number of \textit{Undetected} individuals at a given time is estimated by assuming that the detecting ratio, that is the percentage of detected cases with respect to the total number of positive cases, be computed as
\begin{equation}\label{eq:stima}
\frac{I(t)+H(t)+T(t)}{U(T)+I(t)+H(t)+T(t)}\approx\frac{\Delta I(t)+\Delta H(t)+\Delta T(t)}{\Delta U(t)+\Delta I(t)+\Delta H(t)+\Delta T(t)}=\frac{\text{IFR}}{\text{CFR}(t+d)}.
\end{equation}
Here we have assumed that the variation of the number of total positive individuals in the time window $[t-\Delta t, t+ \Delta t]$ can be approximated by the variation of the resolved case shifted by a confirmation-to-death delay $d=13$ (see \cite{adjCFR}), namely
$$\Delta U(T)+\Delta I(t)+\Delta H(t)+\Delta T(t) \approx \Delta R(t+d)+\Delta E(t+d) = \frac{\Delta E(t+d)}{\text{IFR}}.$$ 
Similarly, we have assumed that the variation of detected positive individuals in the time window $[t-\Delta t, t+ \Delta t]$ can be approximated with the variation of the resolved detected cases shifted by the same time delay:
$$\Delta I(t)+\Delta H(t)+\Delta T(t) \approx \Delta R_D(t+d)+\Delta E(t+d) = \frac{\Delta E(t+d)}{\text{CFR}(t+d)}.$$ 

By using the available data for $I$, $H$, $T$, and $E$, other than the value of $S$ deduced under the assumption of constant total population and the estimate of $\text{CFR}(t)$ given in Eq.~(\ref{eq:CFR}), we estimate the initial conditions for $R$ and $U$ as
$$
R(t) = \left( \frac{1}{\text{IFR}} - 1\right)\, E(t),
$$
$$
U(t) = \left( \frac{\text{CFR}(t+d)}{\text{IFR}} - 1 \right)\, \left( I(t)+H(t)+T(t) \right),
$$
from Eqs.~\eqref{eq:IFR} and \eqref{eq:stima}, respectively.
}

\section{Parameter calibration}\label{sec:calibration}

Model calibration through data fitting is essential to reproduce the past history of the epidemic and to perform short-term forecasts by inferring the epidemiological characteristics of COVID-19. 

Here we use reported isolated, hospitalized, threatened and extinct cases data to estimate the parameters of the proposed {{\tt SUIHTER}} model. In particular, we perform the calibration in two steps. Firstly, we find a set of parameter values using an (ordinary) least squares (LS) estimator. Then, we perform a Bayesian calibration {using a Markov Chain Monte Carlo (MCMC) algorithm, starting from a prior distribution of the parameters} {centered about the LS estimate.} {Calibration of epidemiological models has been already performed in a Bayesian framework, following the pioneering paper by O'Neill and Roberts \cite{oneill}, for several infectious diseases \cite{Cauchemez2825,DORIGATTI20129,lekone}. In the case of COVID-19 epidemic, Bayesian inference has been performed using simpler SIR 	\cite{piazzola2020note,taghizadeh2020uncertainty},  meta-community SEIR-like \cite{Bertuzzo,Flaxman2020,Gatto,Li489,Marzianoe2019617118} and SEIAR \cite{PEIRLINCK2020113410} models,} in this latter case aiming at estimating nine parameters -- including a dynamic, time-dependent contact rate $\beta(t)$ -- during the first outbreak of the COVID-19 epidemic. In addition to model calibration, 
our analysis also provides a numerical assessment of the predictive capability of the model, in forecasting with an adequate advance notice the occurrence of a peak for the most relevant compartments.

System (\ref{eq:suihter}) can be recast in the following general form describing a system of ODEs for a state vector $\mathbf{Y}$ with $n_e$ components (or compartments): find $\mathbf{Y}(t):[t_I,t_F]\to\mathbb{R}^{n_e}$ with
$\mathbf{Y}(t)=[Y_1(t),\ldots,Y_{n_e}(t)]^T$ such that
\begin{eqnarray}
&&\mathbf{Y}^\prime(t)=\mathbf{F}(t,\mathbf{Y}(t);\mathbf{p}(t))\quad t\in(t_I,t_F]\label{pb:1}\\
&&\mathbf{Y}(t_I)=\mathbf{Y}_0.\label{pb:2}
\end{eqnarray}
where $\mathbf{Y}_0 \in \mathbb{R}^{n_e}$ denotes the initial condition at time $t_I$ \REV{evaluated as discussed in Section \ref{sec:initialization}.}
The evolution of the system depends on $n_{par}$ time-dependent parameters, collected into the function $\mathbf{p}(t):(t_I,t_F]\to\mathbb{R}^{n_p}$.

Let us partition the interval $I=[t_I,t_F]$ into $n_{ph}$ phases, corresponding to different epidemic stages due to, e.g., partial restrictions (such as lock-down measures) or different containment rules introduced by the Government or by the local Authorities. Moreover, assume that on each phase, the value of the $n_{par}$ model parameters is constant (but unknown), so that we can introduce the following set of admissible parameters
\begin{equation}\label{eq:constraints}
\mathcal{P}_{ad}=\{
\mathbf{p}(t): \mathbf{p}(t)\vert_{I_k} = 	\mathbf{p}_{k} \in [\mathbf{p}_{L,k}, \mathbf{p}_{U,k}],  \  k=1,\ldots,n_{ph} \}
\end{equation} 
where $\mathbf{p}_{L,k}, \mathbf{p}_{U,k}$ are given constant vectors. For the sake of notation, let us denote by ${\bf p} \in \mathbb{R}^{n_p}$ the vectors of unknown parameters to be estimated, with $n_p = n_{par} n_{ph}$, and  let $\mathbf{Y} = \mathbf{Y}(t,\mathbf{p})$ highlight the dependence of the states on the parameters. Consequently, $\mathcal{P}_{ad}$ is the $n_p$-dimensional hypercube delimited by the constraints \eqref{eq:constraints}. {Additional constraints on the parameters are assumed, by imposing that some of them are constant over all phases}. 

\REV{Let $\Delta t$ be a positive time step, for which we consider $n_{me}$ measurements of $n_{com} = 5 < n_e$ compartments at equally spaced times $t_j = j \Delta t$, $j=1, \ldots, n_{me}$ over the interval $I=[t_I,t_F]$, with $t_1 = t_I+\Delta t$, $t_{n_{me}} = t_F$; in total, we have $n_{com} \times n_{me} = 5 \times n_{me}$ reported data, say $\hat{\mathcal{D}}(t)  = \{
\hat{\bf Y}_{I,H,T,E,R_D} (t_j)
\}_{j=1}^{n_{me}} \in \mathbb{R}^{5 \times n_{me}}$, that is,
\[
\hat{\mathcal{D}}(t)  = \{ (\hat{I}(t_1), \hat{H}(t_1),  \hat{T}(t_1), \hat{E}(t_1), \hat{R}_D(t_1))^T, \ldots, 
(\hat{I}(t_{n_{me}}), \hat{H}(t_{n_{me}})),  \hat{T}(t_{n_{me}}), \hat{E}(t_{n_{me}}), \hat{R}_D(t_{n_{me}}) )^T\}.
\]
}

The first stage of the calibration process is then performed by seeking a LS estimate of the parameters vector, given by the solution of the following minimization problem,
\begin{eqnarray}\label{eq:minimization}
\hat{\mathbf{p}}=\text{arg} \min_{\mathbf{p}\in\mathcal{P}_{ad}} \{\mathcal{J}(\mathbf{p})\} 
\end{eqnarray}
where  
\begin{equation}
\mathcal{J}(\mathbf{p}):=
\sum_{j=1}^{n_{me}}\sum_{k=\{I,H,T,E,R_D\}}  \alpha_k(t_j) \|\mathbf{Y}_k(t_j, {\bf p})-\hat{\mathbf{Y}}_{k}(t_j)\|^2 _2
\end{equation}
being $\mathbf{Y}(t_j)$ the solution of 
\eqref{pb:1}-\eqref{pb:2} evaluated at a certain given  instant $t_j$, $j=1, \ldots, n_{me}$ and $\|\cdot\|_2$ the usual Euclidean vector norm. Here, we denote by ${\bf Y}_k$, $k=\{I,H,T,E,R_D\}$ the components of the vector ${\bf Y}$ corresponding to the compartments $I,H,T,E$, and $R_D$, respectively, and by
$\mathcal{D}(t,{\bf p}) = \{ {\bf Y}_k(t,{\bf p}), \ \  k=\{I,H,T,E,R_D\} \}$ the model outcome used for its calibration. 
For a balanced distribution of the error across the different compartments, whose amplitudes vary along time, the dynamical weight coefficients are defined as $\alpha_k(t_j)=1/\hat{{\bf Y}}_k(t_j)$.

We considered the official epidemiological data supplied daily by
the {Italian Civil Protection}, hereafter called ``raw data'' and freely available at
\url{https://github.com/pcm-dpc/COVID-19}, \cite{PCM-DPC}. The accuracy of these data is highly questioned, in particular concerning the estimate of the total number of infection (strongly dependent on the daily screening
effort). \REV{The $n_{com} = 5$ time series selected for model calibration (\textit{Isolated, Hospitalized, Threatened, Extincts} and \textit{Recovered from detected})}  are those considered more reliable among the data daily supplied by the Italian authorities. One of the key features of the proposed {\tt SUIHTER} model is indeed the one-to-one correspondence of the compartments with the categories for which reliable data, as the ones provided on a daily-basis by the Italian Civil Protection, are available \cite{PCM-DPC}.

When $n_{ph}$ phases are considered,  equation~\eqref{eq:minimization} leads to the optimization {of $n_p= 14 n_{ph}$ parameters in total. Namely,  for each phase of the epidemic, we have the $14$ parameters given by $[\beta_U, \beta_I, \beta_H, \omega_I, \omega_H, \delta, \rho_U, \rho_I, \rho_H, \theta_H, \theta_T, \gamma_I, \gamma_H, \gamma_T]$. 

Unfortunately, so many parameters make the calibration process problematic. In what follows, we calibrate our model under the following simplifying assumptions:}
\begin{itemize}
    \REV{\item $\beta_I$ and $\beta_H$ are set to zero, by assuming that the infection only occurs through a contact between a Susceptible individual and an Undetected infected individual;}
    \REV{\item $\theta_H$ is set to zero as we assume that a Hospitalized individual can return back to home only once he has recovered, since this parameters may be difficult to estimate in the absence of specific data on the $H$ to $I$ flux.;}
    \REV{\item $\gamma_H$ are set to zero, by assuming that when a Hospitalized individual is in life-threatening conditions, he/she is moved to ICU;} 
    \item $\delta$, $\rho_U$, $\rho_I$, $\rho_H$, $\gamma_I$, $\theta_T \in \mathbb{R}$  {are constant on $[t_I,t_F]$.}
\end{itemize}
{With these restrictions,} the total number of parameters to be calibrated is reduced to {$4n_{ph}+6$}.

The first stage of the calibration process has been performed by solving the minimization problem \eqref{eq:minimization} numerically.  We have used a parallel version of the limited memory  Broyden-Fletcher-Goldfarb-Shanno algorithm with box constraints (L-BFGS-B), see \cite{L-BFGS-B} for details. 	 

 The second stage of the calibration process aims at quantifying uncertainties and has  been carried out employing a Bayesian framework, since  the latter provides probability densities of the input parameters that can be propagated through the model. 
 
Bayesian inference allows us to construct a probability distribution function (PDF) for the unknown parameters merging prior information  and  available data, these latter entering in the expression of the likelihood function. \REV{At this stage, in order to account for the uncertainty on the initial conditions, we extend the set of parameters to be estimated to $\bar {\bf p}=({\bf p}, {\bf q})$ where ${\bf q}=(U(t_I), R(t_I))$ collects the initial conditions for the \textit{Undetected} and \textit{Recovered} compartments whose values are not available from the data and can only be estimated.} The posterior PDF can then be obtained through the Bayes theorem on conditional probabilities. For the case at hand, we quantify the likelihood of the parameter vector $\bar{\bf p}$ and model outcome \REV{$\mathcal{D}(t,\bar{\bf p})$}  in correlation to the reported cases $\hat{\mathcal{D}}(t)$ as
\[
\pi (\hat{\mathcal{D}}(t) \mid \bar{\bf p} ) \sim N( \mathcal{D}(t,\bar{\bf p}) , \sigma^2 {\bf I})
\]
where ${\bf I} \in \mathbb{R}^{5 \times 5}$ is the identity matrix and the (unknown) variance $\sigma^2$ is assumed to be constant for each compartment. 

\REV{
  Using Bayes’ theorem, we obtain the posterior distribution of the parameters $\bar{\bf p}$ accounting for the prior knowledge on the parameters and the reported cases, as
  \[
 \pi (\bar{\bf p} \mid \hat{\mathcal{D}}(t)) = \frac{ \pi (\hat{\mathcal{D}}(t) \mid \bar{\bf p} ) \pi(\bar{\bf p}) }
{\pi (\hat{\mathcal{D}}(t)) }  =
\frac{ \pi (\hat{\mathcal{D}}(t) \mid \bar{\bf p} ) \pi(\bar{\bf p}) }
{ \int _{\mathcal{P}}   \pi (\hat{\mathcal{D}}(t) \mid \bar{\bf p} ) \pi(\bar{\bf p}) d\bar{\bf p} }, 
 \]
where $\pi(\bar{\bf p})$ denotes the (uniform) prior distribution for the parameters. Here, we assume that the prior PDF for the model parameters $\bf p$ is centered at the LS estimate $\hat{\bf p}$ obtained during the former calibration stage, on a range $[0.9 \hat{\bf p}, 1.1 \hat{\bf p}]$, while priors of the initial values $\bf q$ is centered around an estimate $\hat{\bf q}$ obtained based on the IFR, \textit{infection fatality ratio} (see Eq.~\eqref{eq:IFR}), on a range $[0.7 \hat{\bf q}, 1.3 \hat{\bf q}]$. The larger relative amplitude of the latter prior interval reflects the higher uncertainty on the initial value for the \textit{Recovered} and \textit{Undetected} compartments.}

An alternative, more common and rigorous procedure, would require to specify informative priors for the parameters, starting from key epidemiological features, as done, e.g., in \cite{Gatto}. However, given the large numbers of parameters to be estimated -- some of which do not find explicit counterparts in epidemiological literature -- we have assumed uniform priors, centered about the LS estimates, as a practical shortcut to overcome the difficulty in specifying the prior distribution. In terms of predictive capability of the model, numerical results provided in Section \ref{sec:results} allows us to assess the proposed approach.

Since we cannot obtain the posterior distribution over the model parameters ${\bf p}$ analytically, we adopt approximate-inference techniques based on Monte Carlo (MC) methods, which aim at generating a sequence of random samples from a Markov chain whose distribution approaches the posterior distribution asymptotically, whence the name of Markov chain Monte Carlo (MCMC) \cite{robert2013monte}.  In particular, {we have used the} delayed rejection adaptive Metropolis (DRAM) algorithm \REV{\cite{DRAM}} implemented in \verb+pymcmcstat+, see \cite{miles2019parameter} for the details. The first $500\, 000$ samples of the chain serve to tune the sampler and are later discarded (burn-in period). We use the next $500\, 000$ samples to approximate the posterior distribution for the parameters $\bar{\bf p}$.

From the generated chains, we draw $N_{MC}$ samples of the parameters $\bar{\mathbf{p}}_1, \ldots, \bar{\mathbf{p}}_{N_{MC}}$  that we use to perform forward propagation of uncertainty through the model, and to compute predictive envelopes of the {{\tt SUIHTER}} model compartments (or predictive distributions).

We report the MC samples of the trajectories on the time interval $(t_I, t_{for}]$, including a forecast window $(t_F, t_{for}]$ that extends beyond the time window $(t_I, t_F]$ where data have been reported, to assess the predictive capability of the model.

\section{Results and discussion}\label{sec:results}
In this  section  we  present {three batteries} of numerical results assessing the forecasting capabilities of the {\tt SUIHTER} model.
Our analysis focuses on the second wave of the epidemic that started at the end of the Summer of 2020 and, at the time of this writing, is still affecting Italy. 
{In Section~\ref{sec:secondwave}, we present the simulation of the second wave obtained with the {{\tt SUIHTER}} model using for its calibration all the data between August 20, 2020 and December 31, 2020. By limiting the time range of the data used for the calibration, we also investigate the model capability in forecasting the peaks of the different compartments (see Section \ref{sec:peaks}).} 

\REV{Our results at the national level for the second outbreak have been obtained by initializing  the \textit{Isolated}, \textit{Hospitalized}, \textit{Threatened} and \textit{Extinct} compartments with the data provided by the Dipartimento della Protezione Civile \cite{PCM-DPC} at August 20, 2020, namely $I=15\,063$, $H=883$, $T=68$ and $E=35\,418$. The initial value for the \textit{Undetected} and \textit{Recovered} compartments are estimated using the strategy based on IFR and time-dependent CFR introduced in Section \ref{sec:initialization}, resulting in the values $U=12\,274$ and $R=2\,916\,082$, respectively. Finally, the initial condition for the \textit{Susceptible} compartments is given by $S=N-I-U-H-T-E-R=57\,504\,185$.}  
{Note that this would imply that, by the end of the first wave, around $4.8\%$ of the Italian population had been infected. A serosurvey organized by ISTAT and ISS had estimated that $2.5\%$ of the Italian population had been infected \cite{seroItaly,seroItaly2}; the survey however had a low compliance, so that its results may be biased. A corresponding survey in Spain \cite{seroSpain} with a much higher compliance rate estimated seropositivity to $4.6\%$ or $5\%$, depending on the methodology used for the seroprevalence analysis. Using an ensemble model calibrated over several countries \cite{odriscoll} estimate the proportion infected in Italy at September 1 around 4.5\%. Using instead a dynamical model calibrated over detailed data, \cite{Marzianoe2019617118} estimate the proportion infected in Italy at September 30 of 4.78\%. Thus, the value of \textit{Recovered} cases obtained for August 20 looks rather realistic.}

{
\subsection{Simulation of the second epidemic wave}\label{sec:secondwave}
The {{\tt SUIHTER}} model has been used to simulate the second epidemic outbreak, starting from August 20, 2020 until December 31, 2020. 
The different phases in which the parameters can take different values have been identified according to the occurrence of some critical events:
\begin{itemize}
    \item September 24, 2020: all schools at the national level reopened after the summer (and spring lockdown) closure {(schools calendars vary by grades and by region level in Italy)};
    \item October 8, 2020: new rules imposing the mandatory use of masks in all  locations (either indoor or outdoor) accessible to public;
    \item October 26, 2020: confinement rules including distance learning for most secondary schools, limitations on the activity of shops, bars and restaurants, strong limitation of sport and leisure activities\footnote{DPCM October 24, 2020, \url{http://www.governo.it/sites/new.governo.it/files/DPCM_20201024.pdf}};
    \item November 6, 2020: stricter confinement rules including distance learning from 9th grade, further restrictions on commercial activities, limitations on the circulation outside the own municipality (for some Italian regions, classified as \textit{red} regions)\footnote{DPCM November 4, 2020, \url{https://www.gazzettaufficiale.it/eli/gu/2020/11/04/275/so/41/sg/pdf}};
    \item November 15, 2020: additional confinement rules as more regions turned to \textit{red} color \footnote{\url{http://www.salute.gov.it/imgs/C_17_notizie_5171_0_file.pdf}};
    \item November 19, 2020: additional confinement rules as more regions turned to \textit{red} color\footnote{\url{http://www.regione.abruzzo.it/system/files/atti-presidenziali/ordinanze/2020/ordinanza-n-102.pdf}};
    \item November 29, 2020: relaxation of confinement rules in some regions turned to \textit{orange} color\footnote{\url{http://www.salute.gov.it/imgs/C_17_notizie_5197_0_file.pdf}};
    \item December 6, 2020: relaxation of confinement rules in some regions turned to \textit{yellow} color\footnote{\url{https://www.gazzettaufficiale.it/eli/id/2020/12/05/20A06781/sg}};
    \item December 18: stricter confinement rules are introduced for Christmas holidays\footnote{Decree Law December 18, 2020, n. 172, \url{https://www.gazzettaufficiale.it/eli/id/2020/12/18/20G00196/sg}}.
\end{itemize}
By considering a time lag of 4 days (to account for the incubation period) \cite{ganyani2020estimating}, the corresponding phases on which the model parameters are defined and possibly changing) are:
\begin{itemize}
    \item Phase 1: August 20, 2020 - September 28, 2020;
    \item Phase 2: September 29, 2020 - October 11, 2020;
    \item Phase 3: October 12, 2020 - October 29, 2020;
    \item Phase 4: October 30, 2020 - November 9, 2020;
    \item Phase 5: November 10, 2020 - November 18, 2020;
    \item Phase 6: November 19, 2020 - November 23, 2020;
    \item Phase 7: November 24, 2020 - December 3, 2020;
    \item Phase 8: December 4, 2020 - December 10, 2020;
    \item \REV{Phase 9: December 11, 2020 - December 22, 2020;}
    \item \REV{Phase 10: December 23, 2020 - December 31, 2020.}
\end{itemize}
As mentioned in Section \ref{sec:calibration}, the compartments employed for calibration are only those with more reliable data, namely \textit{Isolated} (I), \textit{Hospitalized} (H), \textit{Threatened} (T), \textit{Extinct} (E) and \textit{Recovered from detected} individuals.

We performed the model calibration by employing the MCMC parameter estimation procedure described in Section \ref{sec:calibration}, over the 10 phases using the data over the full time range from August 20, 2020 to December 31, 2020. The simulations were run for the subsequent 15 days beyond the date associated to the last set of data used for the calibration forecasting the evolution of the epidemic until January 15, 2021. For the new additional phase the values of the parameters are obtained by linearly extrapolating the two (constant) values of the corresponding parameter of the last two phases, located at the final day of each phase, namely phases 9 and 10.

In Figure \ref{fig:calibrationMCMCall}, we report the expected values for the time evolution of the 7 compartments of the {\tt SUIHTER} model as well as the time evolution of additional compartment of the \textit{Daily new positive}, {which corresponds to $\delta \, U(t) $,} and the corresponding $95\%$ prediction intervals obtained by propagating input uncertainties through the model. 

\begin{figure}[t!]
    \centering
        \includegraphics[width=\textwidth]{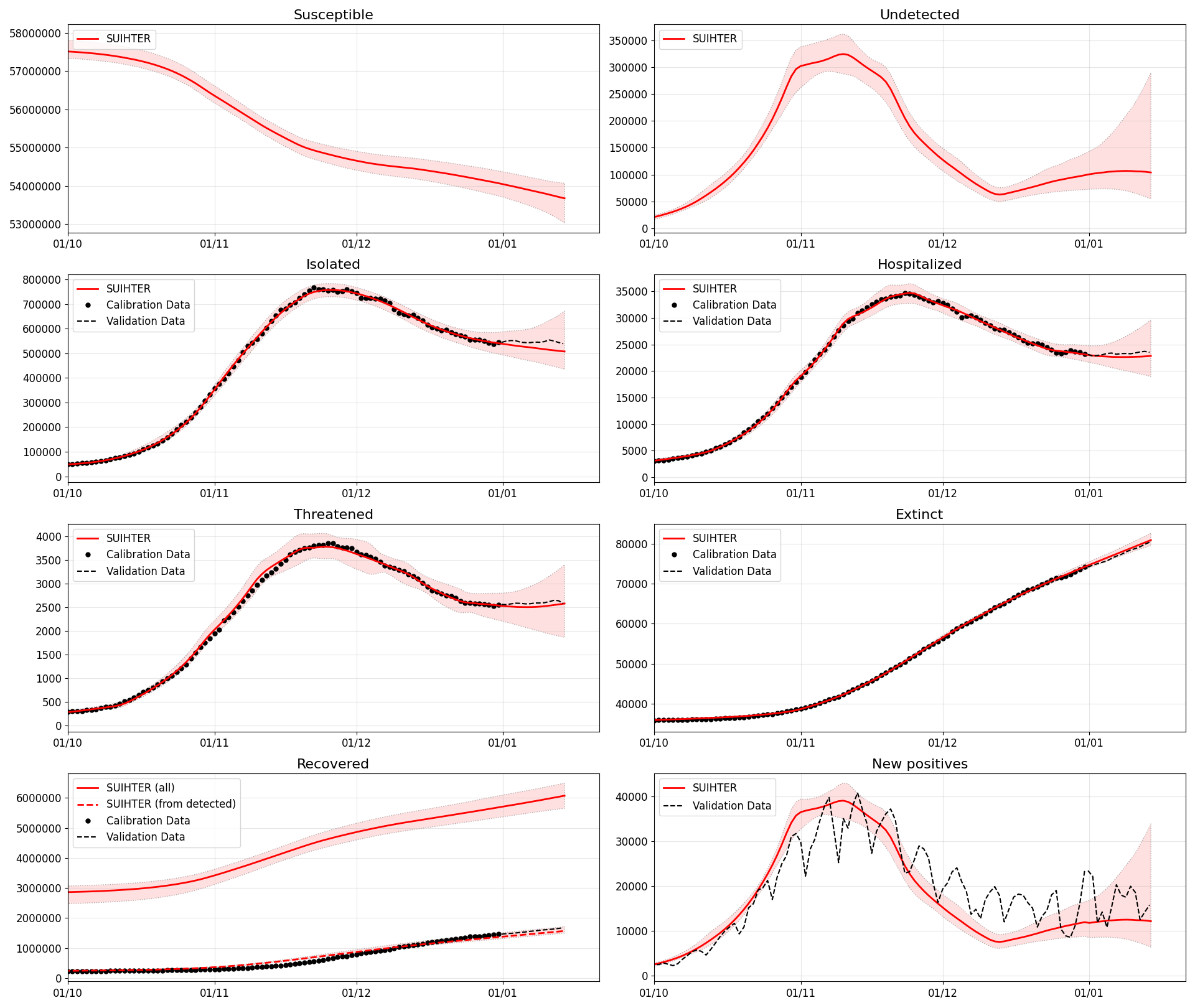}
        \caption{\REV{Expected values (solid lines) and 95\% prediction intervals (shaded areas) for the 7 compartments of the {\tt SUIHTER} model plus the additional \textit{Daily new positives} compartment. The data are indicated with black dots (in the calibration phase) and with a dashed line in the validation phase}} \label{fig:calibrationMCMCall}
\end{figure}

\REV{We notice that the calibrated compartments (\textit{Isolated}, \textit{Hospitalized}, \textit{Threatened}, \textit{Extinct} and \textit{Recovered from detected}) accurately fit the corresponding data time series in the calibration phase. For all the compartment the 15-day forecast also indicates the capability of the model in predicting the evolution of the epidemic at the national level.} 

Moreover, the time history of the \textit{Daily new positives} is also in reasonable agreement with the data, {proving that} the model is able to capture the main dynamics of the system also for quantities that are not directly driven by the data calibration.} \REV{Our calibration indicates that from August 20, 2020 to January 15, 2021, $4\, 056\, 118 \pm  284\, 533$ individuals have been infected, of which $48.0 \% \pm 2.0 \%$ have been detected. In addition, we estimate a posteriori, i.e. by using the outputs of the simulation, that the IFR during the same period is $1.3\% \pm 0.1 \%$. This latter figure is compatible with the IFR estimated at $1.2\%$ for Italy by using estimates by age reported in \REV{\cite{imperialreport}} while somewhat higher than the estimate shown in \cite{odriscoll}.} We also observe that our calculated estimates are likely to be underestimated as the second outbreak is still ongoing at the present time and compartments of isolated and extinct individuals become populated at different time scales.

\begin{table}[t]
    \centering
    \footnotesize
    \begin{tabular}{|c|c c|}
    \hline
    & Mean & Std Dev \\
    \hline
$\delta$    &   0.11965 &   0.00587 \\
$\gamma_I$  &   3.79e-5 &   2.20e-6 \\
$\rho_U$    &	0.12356 &   0.00635 \\
$\rho_I$    &	0.02413 &   0.00139 \\
$\rho_H$    &	0.06680 &   0.00270 \\
$\theta_T$  &	0.05009 &   0.00279 \\
$U(t_I)$    & 12\,429 & 1\,762\\
$R(t_I)$    & 2\,727\,163 &  150\,286\\
\hline
    \end{tabular}
    \vspace{-1mm}
    \caption{\REV{Mean values and standard deviations of constant parameters and $U$ and $R$ initial values}}
    \label{tab:constantpar}
\end{table}

\begin{table}[t]
    \centering
    \footnotesize
    \begin{tabular}{|c|cc|cc|cc|cc|cc|}
\hline
& \multicolumn{2}{|c|}{$\beta_U$}
& \multicolumn{2}{|c|}{$\omega_I$}
& \multicolumn{2}{|c|}{$\omega_H$}
& \multicolumn{2}{|c|}{$\gamma_T$}
& \multicolumn{2}{|c|}{$\mathcal{R}_0$}\\
Phase & Mean & Std Dev & Mean & Std Dev & Mean & Std Dev & Mean & Std Dev & Mean & Std Dev \\
\hline
1  & 0.26440 & 0.008960 & 0.00588 & 0.000320  & 0.01328 & 0.000762  & 0.07614 &  0.004416  & 1.088 & 0.0263 \\
2  & 0.36495 & 0.016058 & 0.00772 & 0.000434  & 0.01916 & 0.001068  & 0.12516 &  0.007226  & 1.508 & 0.0681 \\
3  & 0.34504 & 0.011771 & 0.00934 & 0.000466  & 0.02227 & 0.001181  & 0.08822 &  0.004941  & 1.426 & 0.0489 \\
4  & 0.27485 & 0.013013 & 0.00691 & 0.000346  & 0.02634 & 0.001350  & 0.15562 &  0.008458  & 1.134 & 0.0533 \\
5  & 0.24242 & 0.013460 & 0.00494 & 0.000260  & 0.02577 & 0.001333  & 0.16721 &  0.009086  & 1.003 & 0.0553 \\
6  & 0.17803 & 0.010303 & 0.00423 & 0.000239  & 0.02683 & 0.001456  & 0.19130 &  0.010853  & 0.736 & 0.0426 \\
7  & 0.20978 & 0.012137 & 0.00341 & 0.000189  & 0.02624 & 0.001315  & 0.19014 &  0.010244  & 0.869 & 0.0498 \\
8  & 0.19257 & 0.011057 & 0.00313 & 0.000173  & 0.02503 & 0.001314  & 0.18750 &  0.009730  & 0.797 & 0.0458 \\
9  & 0.30576 & 0.016751 & 0.00309 & 0.000169  & 0.02448 & 0.001294  & 0.19193 &  0.010515  & 1.263 & 0.0700 \\
10 & 0.29610 & 0.017241 & 0.00351 & 0.000194  & 0.02491 & 0.001360  & 0.18702 &  0.010406  & 1.224 & 0.0706 \\
\hline
    \end{tabular}
    \vspace{-1mm}
    \caption{\REV{Mean values and standard deviations of the parameters that changes over the phases and the corresponding $\mathcal{R}_0$}}
    \label{tab:variableparl}
\end{table}

The mean values and the standard deviations computed by the MCMC calibration are reported in Table \ref{tab:constantpar} for the parameters that are constant over the simulation \REV{and for the initial values of \textit{Undetected} and \textit{Recovered}, while} in Table \ref{tab:variableparl} we report the parameters that are free to change in each phase. The former parameters and time dependent functions represent rates that can be used to interpret the dynamics of the second Italian outbreak. For example, large values of $\beta_U$ indicate sustained transmission rates at the corresponding phases. Values of healing rates \REV{$\rho_U$, $\rho_I$ and $\rho_H$ are proportional to the probability of healing for individuals in the compartments U, I and H}, but are inversely proportional to the corresponding average time of healing; the rate $\rho_I$ also incorporates the healing on isolated individuals who are however asymptomatic. {To better understand the role of the parameters, note that if they were constant, $\frac{\rho_I}{\rho_I + \omega_I + \gamma_I}$ would represent the probability for an isolated individual to recover without being hospitalized, and similarly $\frac{\rho_H}{\rho_H + \omega_H + \gamma_H}$ represents the probability for a hospitalized individual to recover without being transferred to ICUs. \REV{In the same way, $\frac{\gamma_T}{\gamma_T + \theta_T }$ represents the probability of dying for an individual in ICUs}, and $\frac{\delta}{\delta+\rho_U }$ represents the probability that an infected individual is detected.}

{Finally, Table~\ref{tab:variableparl} also reports the value of the basic reproduction number $\mathcal{R}_0$ calculated as in Eq.\eqref{eq:r0} for the {\tt SUIHTER} model. The calculation uses the model parameters reported in Tables~\ref{tab:constantpar} and~\ref{tab:variableparl} (columns $1-4$).} 
{Note that the estimates of the standard deviations are strongly influenced by the choice of the prior in the interval centered about the values of the model parameters obtained by the least squares procedure $\pm 10\%$. They should mainly be judged in relative terms.}

{We observe that the value of $\mathcal R_0$ obtained by the calibration reflects the full reopening of educational activities and work restart after holidays, as well as the public health measures and restrictions later introduced by authorities to contain the second epidemic outbreak. In particular, the rise of $\mathcal R_0$ in Phases $2$ and $3$ follows the full schools reopening and restart of working activities from mid September, and probably accounts for seasonality effects too. Restrictions on mobility, schools, businesses and partial lock-downs were introduced in late October at regional and national levels, as reflected by the decrease of $\mathcal R_0$ \REV{from Phase~$5$ to $6$}, when $\mathcal R_0$ became smaller than one. Partial reopening and easing restrictions were gradually introduced in some regions and at the national level from late November, as the new increment of $\mathcal R_0$ from Phase~$9$ indicates.}

\subsubsection{{Simulating the second outbreak for Italian regions}}
{The results obtained simulating the epidemic at the national scale can indeed hide specific local outbreaks. The {\tt SUIHTER} model can also simulate the evolution of the epidemic for everyone of the $20$ Italian Regions for which the same data time series as those used for the national calibration are available. Unfortunately, this is not true for the finer geographical level (the $107$ provinces) since only the number of total cases from the beginning of the epidemic is provided.}

\begin{figure}[t!]
    \centering
    Lombardy       \\ \includegraphics[width=\textwidth]{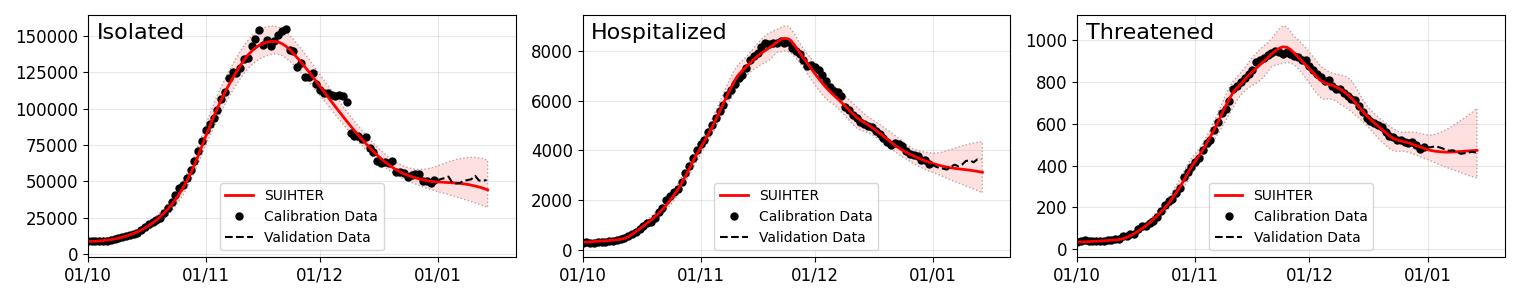} \\[-1mm]
    Veneto       \\ \includegraphics[width=\textwidth]{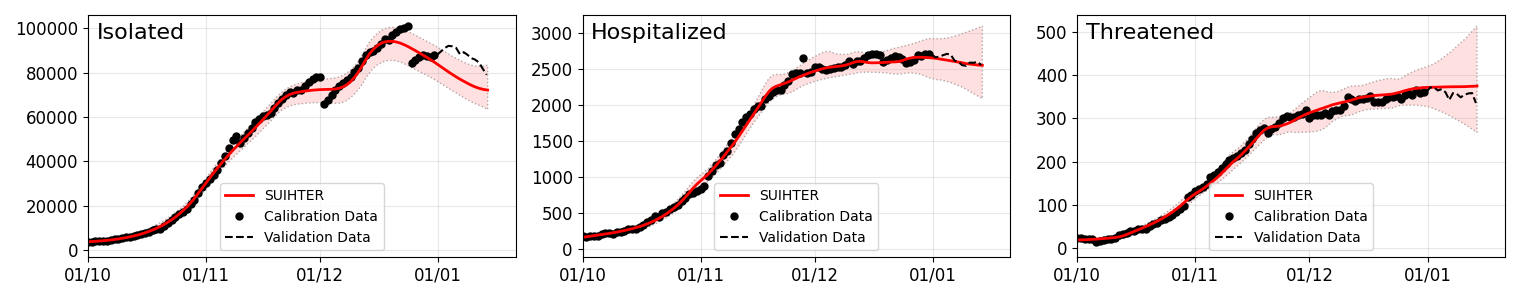} \\[-1mm]
    Emilia-Romagna \\ \includegraphics[clip,width=\textwidth]{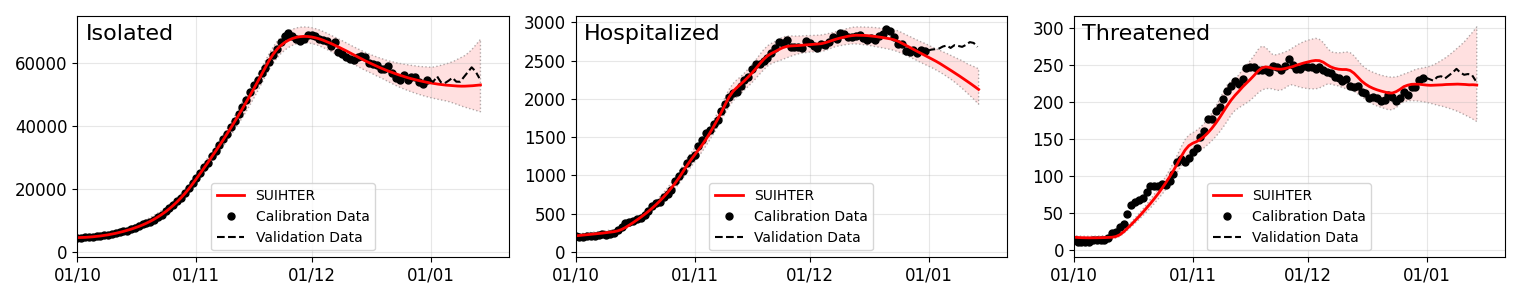} \\[-1mm]
    Lazio          \\   \includegraphics[clip,width=\textwidth]{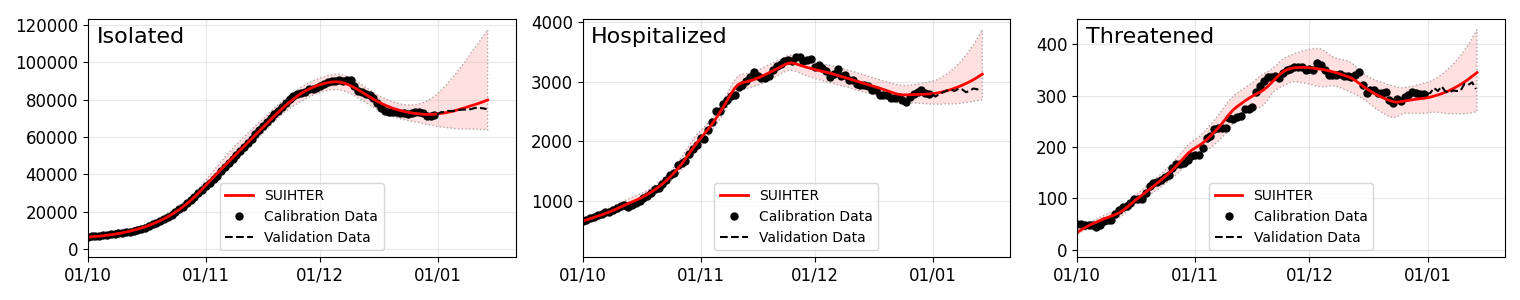}\\[-1mm]
    Campania     \\   \includegraphics[clip,width=\textwidth]{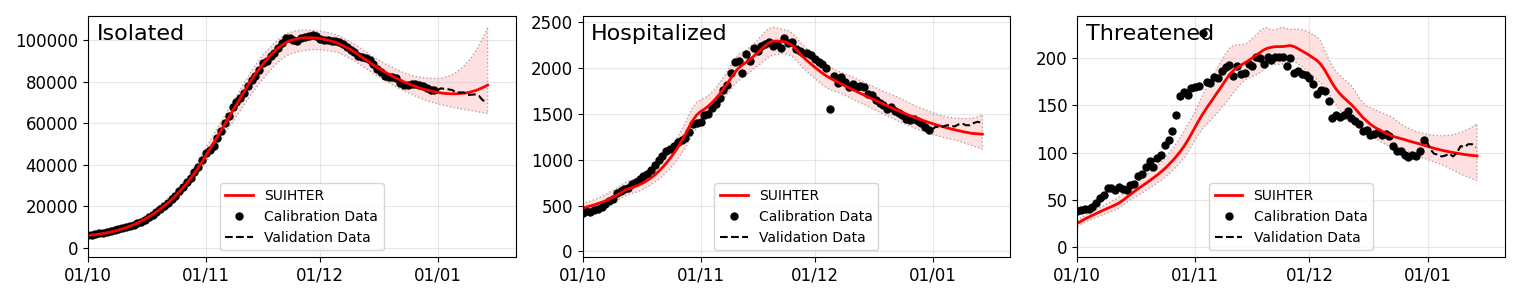}\\[-1mm]
    Sicily          \\   \includegraphics[clip,width=\textwidth]{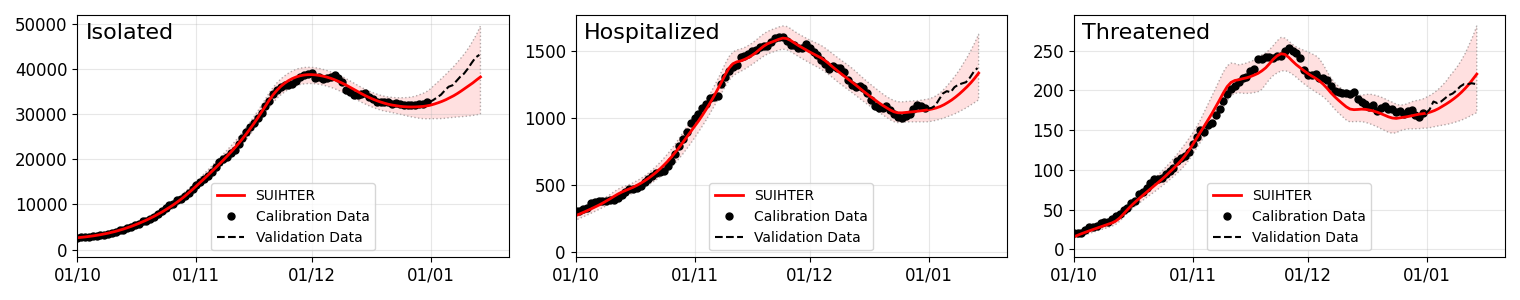}
        \caption{{Expected values (solid lines) and 95\% prediction intervals (shaded areas) for the  \textit{Isolated}, \textit{Hospitalized} and \textit{Threatened} compartments, from left to right, in the six larger Italian regions.}}\label{fig:calibrationMCMCregions}
\end{figure}

\clearpage

\REV{Following the same initialization and calibration strategies adopted at the national level, we have carried out the simulation of the second epidemic outbreak in the six larger Italian regions, namely Lombardy, Veneto, Emilia-Romagna, Lazio, Campania and Sicily. In Figure~\ref{fig:calibrationMCMCregions}, the expected value for the time evolution of the three infectious compartments used for the calibration and the corresponding $95\%$ prediction intervals are reported for the former three regions.} \REV{The calibration has been carried out using the same setting as for the national level, i.e. calibrating the model with the data available until December 31, 2020 and then simulating until January 15, 2021.
The results obtained by numerical simulations stand in good agreement with the real data, with few exceptions, namely the \textit{Isolated} compartment in Veneto (where the time series is clearly affected by some reporting problems) and the \textit{Hospitalized} compartment in Emilia-Romagna.}

\subsection{Predicting the peaks}\label{sec:peaks}
{
Predicting the peak of an epidemic outbreak is a tremendous challenge for an epidemiological model. 
Yet, the predictive capability of epidemiological models is of paramount importance to inform policymakers about the 
dynamics of the disease and foresee
timing and level of peaks of infected, hospitalized and ICU treated individuals, as well as the potential effects of policy responses.

With the goal of investigating to which extent our {\tt SUIHTER} model is able to predict the occurrence of the epidemic wave peak, we repeated the calibration using the data over limited time ranges.
}

 In particular, we have considered {three} different cases: in \textit{Case 0} we used all the data time histories available until December 3, while in \textit{Cases 1}, \textit{2} and \textit{3}, the data employed for the calibration were limited to November 23, November 18, and November 9, respectively.  {For each case, the simulations were run for the subsequent 30 days beyond the date associated to the last set of data used for the calibration and the linear extrapolation carried out as indicated before.}

\begin{figure}[t]
    \centering
        \includegraphics[width=\textwidth]{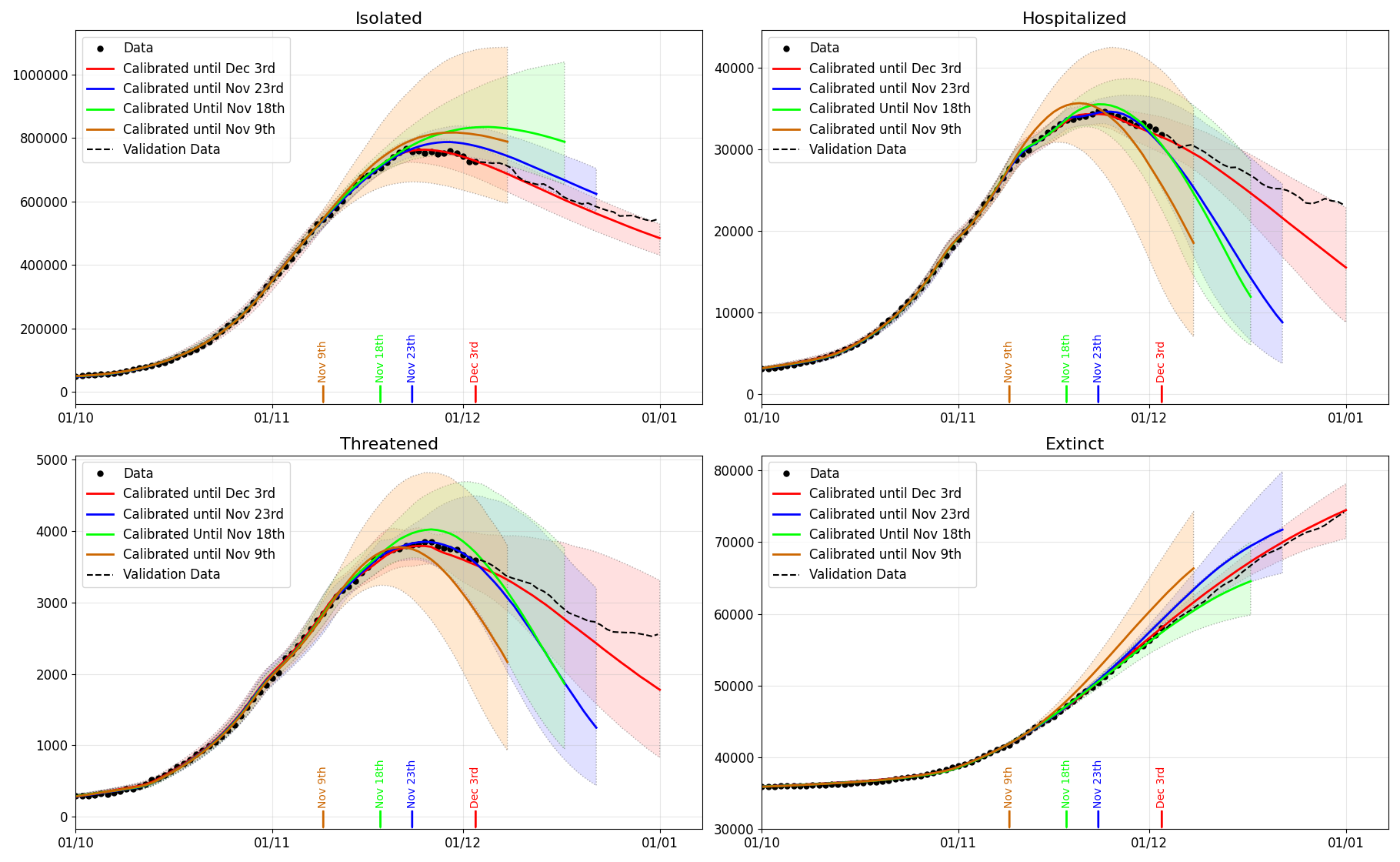}
        \caption{\REV{Peak forecast obtained by the {\tt SUIHTER} model with different data ranges for the  \textit{Isolated}, \textit{Hospitalized}, \textit{Threatened} and \textit{Extinct} compartments}} \label{fig:calibrationMCMC}
\end{figure}

In Figure \ref{fig:calibrationMCMC}, we report the expected value for the time evolution of the {four} compartments used for the calibration, and the $95\%$ prediction intervals obtained by propagating input uncertainties through the model. The accuracy of the forecast, as expected, improves when a richer set of data are employed in the calibration.  Our simulations show the occurrence of a peak for each of the three compartments, not only for \textit{Case 0} in which the time lapse of the data used for the calibration covers the peaks, but also for \textit{Case 1, 2},  and \textit{3}, when the data time-series employed for the calibration are still rising. However, we should remark that if the model is calibrated with a shorter time series, namely available data stop more than 30 days before the peak, the occurrence of the peak cannot be correctly predicted. 

As already noticed, because of the overall complexity of the problem and the limited data available for its calibration, by no means we intend here to certify in rigorous terms the actual values of the future compartments.
However, in spite of the widths of the predictive intervals (which depend, at some extent, on the widths of the chosen prior distributions), we nonetheless observe that the expected values (solid lines in Figure \ref{fig:calibrationMCMC}) carry meaningful prediction capabilities.

\REV{To further assess the accuracy of the prediction, it is interesting to compare this peak forecasting with respect to the actual data, i.e. the day and value that have been reported for the different compartments at the end of November 2020. Moreover, we propose a comparison with the predictions obtained using the two different strategies based on data fitting. The first is based on a simple polynomial fit of degree $2$ on the last recorded $10$ days, while the second is obtained by using a curve registration (see \cite{R10} for an overview on this subject) by exploiting the similarities between the first and second waves.
The registration procedure is performed by first computing the Exponentially Modified Gaussian (EMG) function that best fits the first wave. We denote this function as $w(t), t_0 \le t \le t_1$, with $t_0$ the first day of the recorded data (February 24) and $t_1$ equal to August 1. Then a second minimization problem is solved to compute the time shift and scaling factors to apply to the computed EMG function to best fit the rising portion of the second wave in the time range $[t_k,t_n]$, with $t_k$ coinciding with October 15 and $t_n$ with the last recorder date. Namely, we look for the optimal time shift $\bar{h}$, and the scaling factors $\bar{s}_1$ and $\bar{s}_2$ such that 
$$(\bar{h}, \bar{s}_1, \bar{s}_2) = \argmin_{h, s_1, s_2} \sum_{i=t_k}^{t_n} \left(s_1 \, w(s_2 \, t_i + h) - d_i\right)^2,$$
where $d_i$ is the value of the considered data series at day $t_i$.

For each data series, the fitted EMG function and the optimal values for shift and scaling factors are computed and, in this way  the shape of the first wave can be used to complete the second wave for the different compartments.}

\begin{figure}[t!]
\centering
    \begin{tabular}{cc}
        Isolated (peak day) & Isolated (peak value) \\
        \includegraphics[trim={0 0 0 30},clip,width=0.50\textwidth]{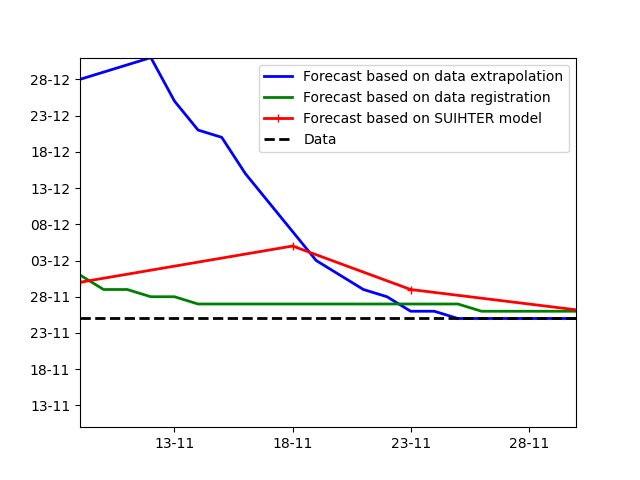}
        &
        \includegraphics[trim={0 0 0 30},clip,width=0.50\textwidth]{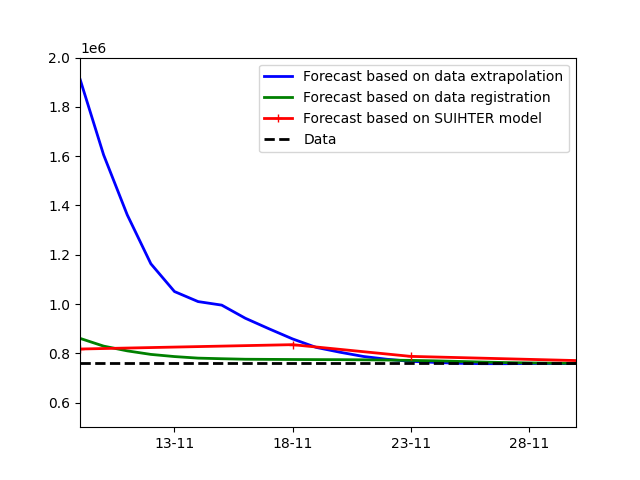}
        \\
        Hospitalized (peak day) & Hospitalized (peak value) \\
        \includegraphics[trim={0 0 0 30},clip,width=0.50\textwidth]{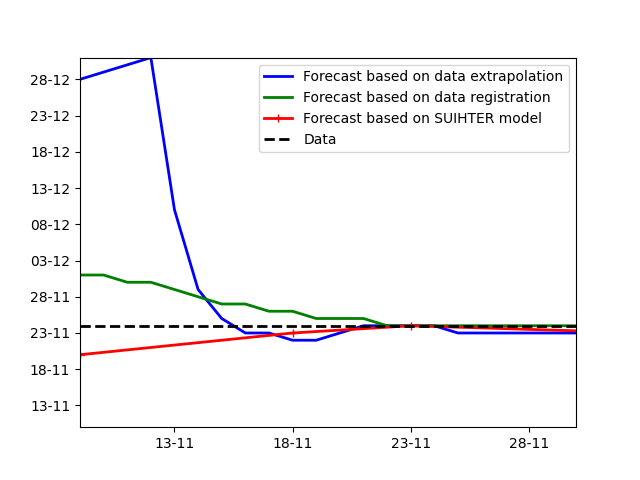}
        &
        \includegraphics[trim={0 0 0 30},clip,width=0.50\textwidth]{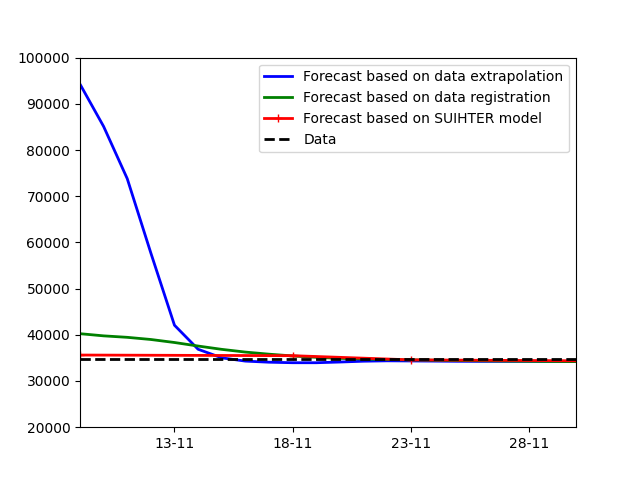}
        \\
        Threatened (peak day) & Threatened (peak value) \\
        \includegraphics[trim={0 0 0 30},clip,width=0.50\textwidth]{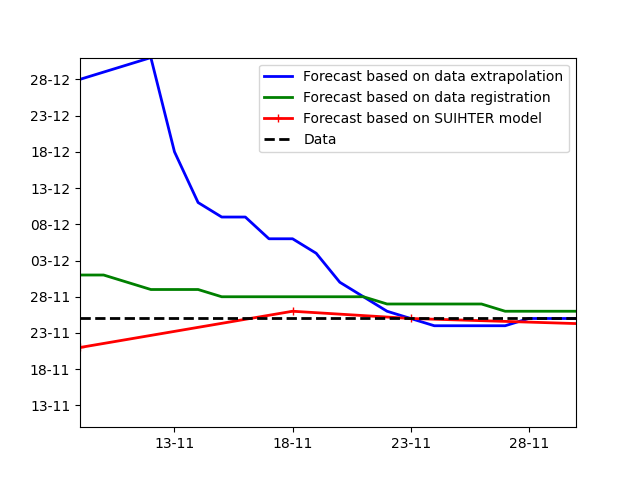}
        &
        \includegraphics[trim={0 0 0 30},clip,width=0.50\textwidth]{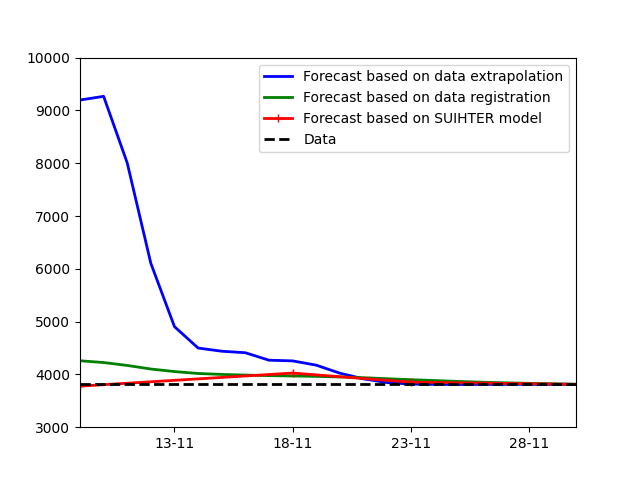}
        \end{tabular}
        \caption{\REV{Peak day (left) and peak value (right) {vs. last used data by day} for the three compartments \textit{Isolated} (top), \textit{Hospitalized} (middle) and \textit{Threatened} (bottom), estimated with data extrapolation, data registration and {\tt SUIHTER} model}} \label{fig:calibrationPeakT}
\end{figure}

A comparison between the peak forecast obtained with the {\tt SUIHTER} model, the quadratic extrapolation (based on the last $10$ days), and the registration approach is displayed in {Figure~ \ref{fig:calibrationPeakT}, for the \textit{Isolated}, \textit{Hospitalized} and \textit{Threatened} compartments.} The curves show how the prediction in terms of day of peak occurrence and peak value changes as far as an increasing number of data are used (the last data day is reported on the horizontal axis). \REV{To minimize the effect of data daily noise, the reference value (in dashed line) is obtained smoothing the data with a Savitzky-Golay polynomial smoothing filters of degree 3  \cite{parolini2021mathematical}.}

\clearpage

By comparing the peak predictions with the day and value of the measured \REV{(smoothed)} data peak (reported with a dashed line in {Figure~\ref{fig:calibrationPeakT}),} we should first remark that the {\tt SUIHTER} prediction largely outperforms  those obtained with polynomial extrapolation. Moreover, even when compared with predictions based on the registration with the first epidemic wave, the {\tt SUIHTER} model is more accurate for most of the considered quantities. When making this comparison, it is worthy noticing that, while  prediction based on the registration strongly depends on the evolution of the different compartment during the first epidemic wave, the predictions based on {\tt SUIHTER} do not require any a-priori knowledge of previous epidemic waves.

\section{Conclusions and model limitations}\label{sec:conclusions} In this paper, we have introduced a new mathematical model, named {{\tt SUIHTER}}, to describe the ongoing pandemic of coronavirus disease 2019 (COVID-19). This epidemiological model is constructed on seven compartments -- susceptible uninfected individuals ($S$), undetected (both asymptomatic and symptomatic) infected ($U$), isolated ($I$), hospitalized ($H$),
threatened ($T$), extinct ($E$) and recovered ($R$) -- and we exploit it to study and analyse the {second Italian outbreak emerged in Fall 2020 and still ongoing}. In particular, our model is suited for calibration against data made available daily by the Italian Civil Protection \cite{PCM-DPC}. On the basis of these data at the national level, our calibration populates {the compartments $I$, $H$, $T$, $E$ \REV{and $R_D$}}, which we purposely use to determine transmission rates, rates of recovery, infection fatality ratio, etc. In particular, {{\tt SUIHTER}} \REV{allows to estimate} the infected, but undetected population, a compartment ($U$) that is crucial for studying and understanding the epidemic, especially considering that large shares of infected individuals went uncounted during the first and even the second outbreaks in Italy. {Moreover, thanks to our approach transmission rates, and thus the basic reproduction number $\mathcal{R}_0$, can be estimated.} Finally, our calibration is made robust by exploiting Bayesian estimation using the Markov Chain Monte Carlo method.

{The {{\tt SUIHTER}} model calibrated at the Italian national level is validated against data {related to} the last part {of the second outbreak.} Comparisons are made against basic statistical models, namely quadratic regression and registration of the first epidemic wave. {The comparison demonstrates} the better accuracy of {{\tt SUIHTER}} for predictive purposes. This is made possible by using extrapolated transmission rates that are calibrated at earlier times through regression models, a feature that allows capturing peaks of the second Italian outbreak correctly, {and enables} using {{\tt SUIHTER}} in a predictive fashion by leveraging data available at the current date. This novel approach attempts to circumvent a common issue {of the use} of epidemiological compartmental models for forecasting \cite{PEIRLINCK2020113410}, that is accurately capturing transmission rates. However, as our approach is based on interpolating values of these transitions rates, the accuracy of their extrapolation and, consequently, their exploitation for prediction within {{\tt SUIHTER}} can only be limited to restricted time windows,} {especially when government interventions and citizen behaviours are changing).
Note that, although the calibration procedure did not make any assumptions about the temporal changes in parameters, the estimates accurately reflect the policy changes: estimates of $\mathcal R_0$ decrease as control measures are tightened and increase when they are relaxed.} \REV{The results of the simulation of the second wave carried out at the national and regional levels showed the capability of the model in predicting accurately the time-evolution in a time frame of 15 days past the data used in the calibration. Longer term predictions the model should account for the possible changes in restriction rules that may occur in the future to supply analyses of different scenarios (as recently done in \cite{parolini2021mathematical} based on the {\tt SUIHTER} model).}

{A further limitation of our approach is that we are currently calibrating the Italian epidemic outbreaks at the national level, that is as a whole, without summing up the different contributions at the level of the $20$ Italian regions for which data are available \cite{PCM-DPC}; we indeed performed the calibration of the \REV{six larger Italian regions}. Populating compartments at the national level by summing up results obtained by tailored calibrations of each Italian region would allow a better capturing of the spatio-temporal heterogeneity of the Italian outbreaks, which reflects different mobility patterns {and density of population}. In this respect, several different approaches have been proposed in literature, see, e.g. \cite{book-chen} and the references therein, ranging from the use of network based models  \cite{Vespignani,Colizza}, to systems of ordinary differential equations on network \cite{Allen,Arino}, {as well as} non-local partial differential equations \cite{Yang}. Among the contributions appeared during the COVID-19 pandemic, we also recall the recent papers \cite{Bertuzzo,Gatto,Li489}, where a meta-community SEIR-like model has been proposed and employed to reproduce the contagion in Italy. Still, calibrating our {{\tt SUIHTER}} at the regional level, and for all the regions, would require a more sophisticated design due to the intrinsic ill-posedness of the inverse problem, especially {when taking mobility patterns into account.} Nevertheless, we plan to better address spatio-temporal heterogeneity of the Italian outbreaks in the future by generalizing our {{\tt SUIHTER}} model to incorporate suitable spatial-multicity mobility terms at the regional level. Even if a more spatially detailed compartment model is surely desirable, to act, for example, at the province level (Italy is comprised of $107$ provinces), at the time being no detailed data for its calibration have been made available. }

{Albeit the {\tt SUIHTER} is namely very sophisticated and involves \REV{$14$} time-dependent parameters and functions to be determined based on available data, we limited our calibration to a subset of the possible control variables, by forcibly setting to zero some parameters that we deemed to be less relevant for the transmission of the epidemic and by assuming as constant over time some other ones. {We also neglected incubation time, and we implicitly assumed that all distributions in the states are exponential, which is far from correct \cite{Ferrettieabb6936}. Still, we believe that} this qualifies as an acceptable compromise among the complexity of the {\tt SUIHTER} model and its calibration procedure, the associated computational costs, and the accuracy of the results. Some of the calibrated parameters assume values that are able to compensate for those parameters prescribed a-priori, even if their interpretation may not result straightforward in explaining the outbreak. In this respect, we plan to assess the robustness of our approach by allowing the calibration of additional parameters. Further, our multi-compartment {\tt SUIHTER} model does not consider stratification of ages groups within the compartments. This is namely an important aspect as some compartments like $H$, $T$ and $E$ are mostly populated by the elderly, while the transmission mechanisms widely differ by age and context of infection (workplace, school, family, etc.) \cite{DiDomenico2020,Marzianoe2019617118}. We also plan to improve {\tt SUIHTER} by considering age stratification within its compartments.}

{Finally, in consideration of the ongoing emergency situation amidst the second Italian outbreak, we believe that our {{\tt SUIHTER}} model is well suited
to be used in a predictive manner to support and motivate public health measures.} To the best of our knowledge, apart from \cite{Cereda20} wherein a SEIRD model is used at the regional level, {{\tt SUIHTER}} stands as one of the first models to analyze the second Italian COVID-19 outbreak and that can readily serve the purpose {of predicting the short-term epidemic trends and perform longer term scenario analyses}. 

\section*{Acknowledgements}
The authors would like to thank Prof. Luca Formaggia for his insightful suggestions and careful reading of the manuscript.

\section*{Funding statement}
This research received no specific grant from any funding agency in the public, commercial, or not-for-profit sectors.


\end{document}